\documentclass[prb,preprint]{revtex4-1} 

\usepackage{amsmath}  % needed for \tfrac, \bmatrix, etc.
\usepackage{amsfonts} % needed for bold Greek, Fraktur, and blackboard bold
\usepackage{graphicx} % needed for figures
\usepackage{epstopdf}
\newcommand{\R}{\mathbb{R}}
\newcommand{\C}{\mathbb{C}}
\newcommand{\N}{\mathbb{N}}

\newcommand{\BW}[1]{}       % put in the dash-1 to switch to black-white mode
\newcommand{\color}[1]{#1}      % put in the dash-1 to switch to color mode
\begin{document}

\title{A New Way of Visualizing Quantum Fields}
\author{Helmut Linde}
\email{Helmut.Linde@gmx.de} % optional
\altaffiliation[permanent address: ]{Am M\"arzenb\"achel 1, 
  69226 Nu\ss loch, Germany} 
\date{\today}

\emph{This is an author-created, un-copyedited version of an article accepted for publication/published in European Journal of Physics. IOP Publishing Ltd is not responsible for any errors or omissions in this version of the manuscript or any version derived from it.  The Version of Record is available online at https://doi.org/10.1088/1361-6404/aaa032.}

\begin{abstract}
Quantum Field Theory (QFT) is the basis of some of the most fundamental theories in modern physics, but it is not an easy subject to learn. In the present article we intend to pave the way from quantum mechanics to QFT for students at early graduate or advanced undergraduate level. More specifically, we propose a new way of visualizing the wave function $\Psi$ of a linear chain of interacting quantum harmonic oscillators, which can be seen as a model for a simple one-dimensional bosonic quantum field. The main idea is to draw randomly chosen classical states of the chain superimposed upon each other and use a \BW{gray}\color{color} scale to represent the value of $\Psi$ at the corresponding coordinates of the quantized system. Our goal is to establish a better intuitive understanding of the mathematical objects underlying quantum field theories and solid state physics. 
\end{abstract}

\maketitle % title page is now complete

%-------------------------------------------------------
\section{Introduction} % 
%-------------------------------------------------------

Quantum Field Theories (QFT) are celebrated for being the framework of the Standard Model and for making predictions which coincide with experimental data to extreme accuracy. Yet QFT is perceived as difficult to learn by many and this seems to be at least partially due to the lack of visualization options which would help to develop an intuition for the theory. In  quantum mechanics (or `first quantization'), for example, a localized particle is associated with a $\delta$-function and a uniformly moving particle with a plane wave, both of which are relatively easy to depict in a graph. Many textbooks show the energy eigenstates of hydrogen atoms, harmonic oscillators or quantum wells as plotted wave functions. A survey of visualization methods in quantum mechanics can be found in Thaller's books \cite{Thaller}$^,$\cite{Thaller2}. In QFT, on the other hand, visualizations of the fundamental concepts seem to be scarce. One prominent exception are the Feynman diagrams, which are an important tool in studying processes of interacting particles. Yet these diagrams are firmly connected to the specific computational method of perturbation theory and they are not meant as a graphical representation of the quantum field itself. Another approach is to compute and plot statistical quantities, like the particle density or the two point correlation function of a many body system. But a lot of information is lost when moving from the full description of its state to these aggregated quantities. Therefore they tend to be more useful to understand macroscopic properties of the system rather than the underlying details. 

The difficulties in visualizing a quantum field obviously stem from the fact that such a system has too many degrees of freedom: Even if the field is simplified to a discrete lattice with $N$ atoms oscillating in only one space dimension each (like, e.g., in the `quantized mattress' model in Zee's book\cite{Zee}), the quantum state is given by a wave function  $\Psi: \R^N \rightarrow \C$, which cannot be plotted in a straight-forward way for any $N > 2$.

Such lattices of coupled quantum oscillators are also studied in several other areas of physics and especially in solid state physics they are applied to model phonons in a crystal.
Johnson and Gutierrez\cite{JohnsonGutierrez} visualize phonon states of a one-dimensional quantum lattice by projecting the probability density of the system to each of the one-dimensional spaces in which the atoms oscillate. 

In the present article we propose a new way of visualizing the state of a bosonic quantum field and apply it to the example of the harmonic chain, which is essentially a discretized, one-dimensional model for a boson field. The target audience for this material are mainly graduate students who completed the courses on quantum mechanics and who are now about to advance towards quantum field theory. We assume that the reader is familiar especially with the treatment of the quantum harmonic oscillator and base changes in Hilbert space via Fourier transformation. 

The article is organized as follows: In Section \ref{SecMonteCarlo} we present a Monte Carlo method for visualizing quantum wave functions in (first quantization) quantum mechanics. Then we introduce a quantum harmonic chain in Section \ref{SecLinHarChain} and we compute its dynamics by diagonalizing its Hamiltonian via a Fourier transformation. In Section \ref{ConnectingDots} we explain how the harmonic chain can be interpreted in the pictures of solid state physics (where each atom in the chain is a particle) and in quantum field theory (where a particle is an excitation of the chain). Setion \ref{SecVisualizationOfChain} contains our main result - a new visualization method for the state of a quantum harmonic chain and the application to several interesting special cases. We conclude with a discussion of our findings and an outlook to future research in Section \ref{SecDiscussion}. 

%-------------------------------------------------------
\section{Monte Carlo plot of wave functions} \label{SecMonteCarlo}
%-------------------------------------------------------

De Aquino et. al. have proposed the following method to visualize the wave function $\Psi: \R^2 \rightarrow \C$ of a single quantum particle in two dimensions\cite{DeAquinoEtAl}: Associate $\R^2$ with the two orthogonal axes of a scatter plot and then draw many dots into this plane, with the location of each dot being randomly chosen according to the probability density function $|\Psi|^2$. The resulting charts show a high density of dots where $|\Psi|^2$ is relatively large and only very few dots where $\Psi$ is close to zero.

In our visualization method we follow a similar Monte Carlo approach, but we impose the following rules:
\begin{enumerate}
\item Choose the locations of the dots randomly according to a uniform probability distribution in a rectangular window within $\R^2$.
\item Represent the value of $\Psi$ at each dot via its color. For energy eigenstates, which can always be written as real-valued functions, such a visualization is possible on a black and white scale. For complex-valued functions a color spectrum can be used. 
\end{enumerate}

\begin{figure}[h!]
\centering
\BW{\includegraphics[width=180mm]{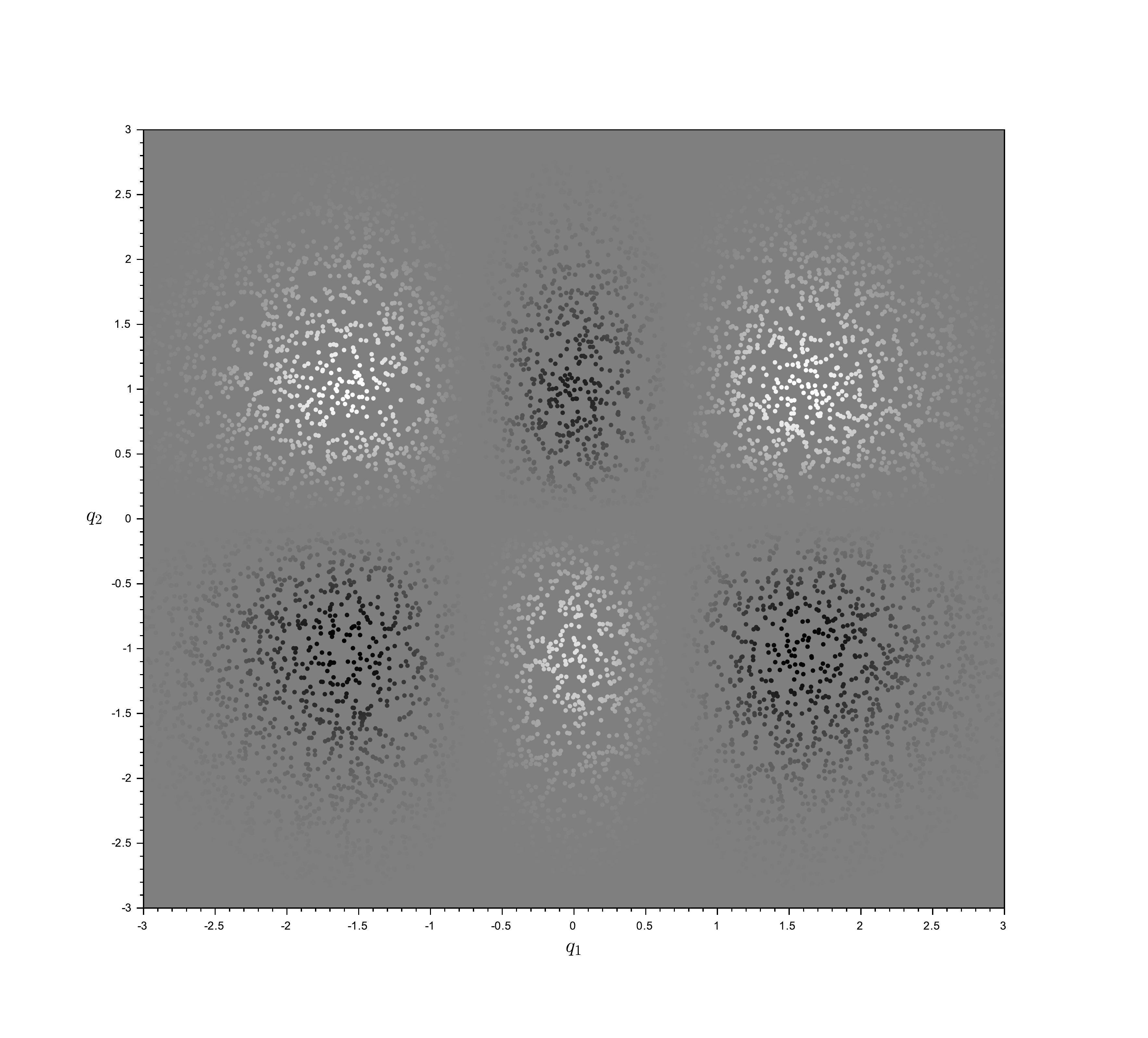}}
\color{\includegraphics[width=180mm]{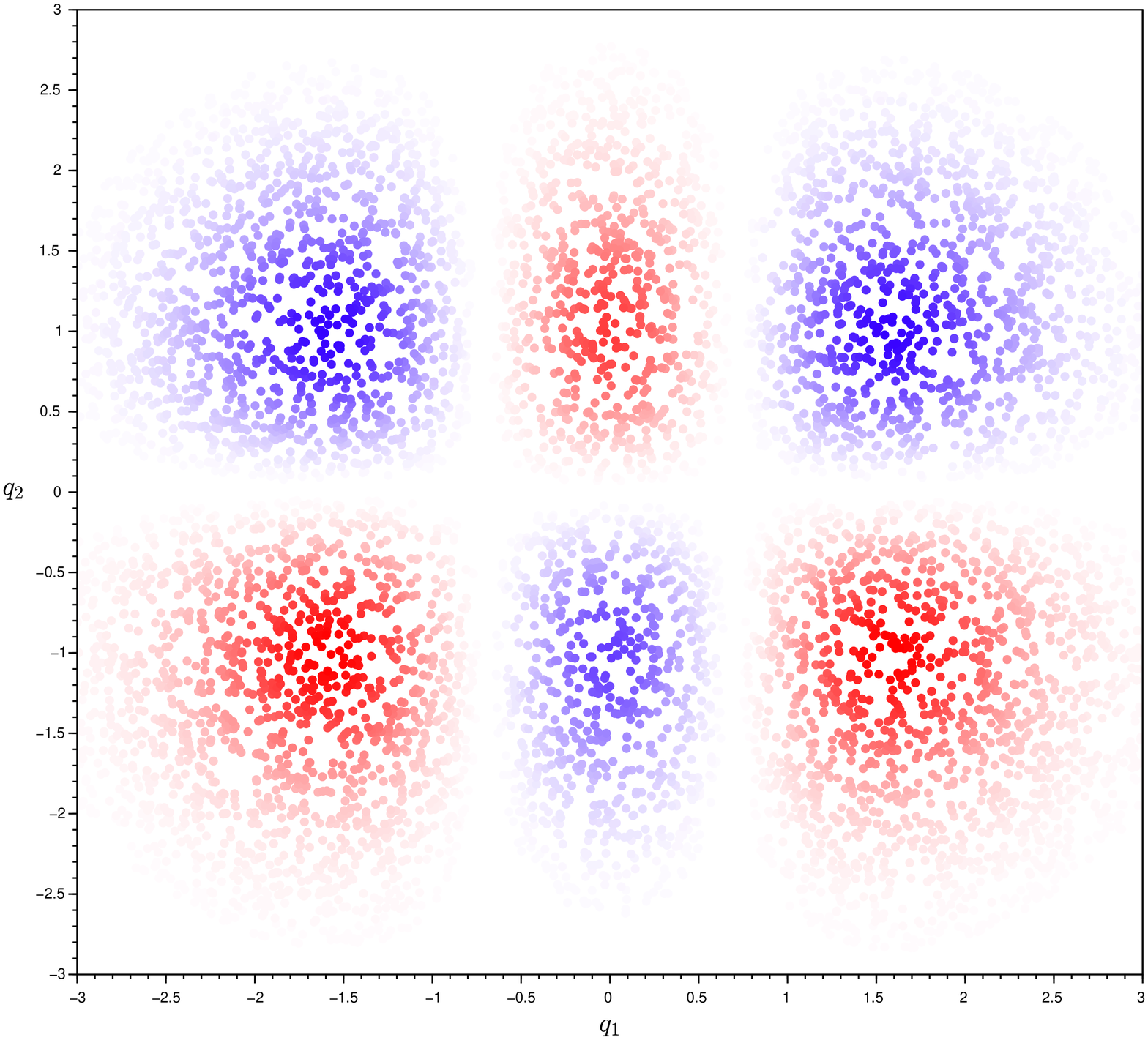}}
\caption{Monte Carlo representation of the two-dimensional harmonic oscillator energy eigenstate $\Psi_{2,1}$. \BW{Bright (dark)}\color{Red (blue)} dots correspond to positive (negative) values of the wave function. Horizontal and vertical nodal lines of the function can be recognized as areas without any visible dots. As expected, the state $\Psi_{2,1}$ has two nodal lines intersecting with the $q_1$ axis and one intersecting with the $q_2$ axis.}
\label{2DHarmonic}
\end{figure}

To start with a simple example from ordinary quantum mechanics, consider a two-dimensional harmonic quantum oscillator with the Hamiltonian 
\begin{equation}
\hat H = \frac{\hat{p_1}^2}{2m}+\frac{\hat{p_2}^2}{2m} + \frac{1}{2} \kappa (\hat{q_1}^2 + \hat{q_2}^2),
\end{equation}
where $q_l$ with $l \in \{1;2\}$ are the position coordinates of the particle with mass $m$, $\hat{p_l} = -i \partial/\partial q_l$ are the momentum operators, and $\kappa$ controls the strength of the attractive potential. Here and in the rest of the article we use units of measure such that $\hbar = 1$. It is well known\cite{Liboff} that the Hamiltonian is separable and the (non-normalized) energy eigenstates of the system can be written as
\begin{equation}
\Psi_{\nu_1,\nu_2} = \prod_{l=1,2}{H_{\nu_l}(\sqrt{m\omega}\,q_l)e^{-m\omega q_l^2/2}},
\end{equation}
where $\omega^2 = \kappa/m$,  $H_\nu$ is the $\nu$-th order Hermite polynomial and $\nu_{1,2} \in \N_0$ are the two quantum numbers which enumerate the energy eigenstates of the system.
As an example, Fig. \ref{2DHarmonic} shows a Monte Carlo plot of the energy state $\Psi_{2,1}$ created according to the two rules above. \BW{Black (white)}\color{Blue (red)} dots represent negative (positive) values of $\Psi_{2,1}$ and dots where $\Psi_{2,1} \approx 0$ blend in with the chart's \BW{grey}\color{white} background\BW{, i.e., the brighter a dot is, the larger is the value of $\Psi_{2,1}$ in the respective point of $\R^2$}. We have omitted a quantitative color scale since the normalization of the wave function is not important. It is easy to recognize the general shape of $\Psi_{2,1}$ and its nodal lines.

Obviously such a Monte Carlo plot is only of limited relevance in two dimensions, since we could have simply colored the whole chart area according to the value of $\Psi_{2,1}$ instead of just picking a few points, or we could have drawn the wave function in a 3D plot. We will see later that such a Monte Carlo visualization actually becomes quite useful when moving beyond two dimensions.

%-------------------------------------------------------
\section{Quantized Linear Harmonic Chain} \label{SecLinHarChain}
%-------------------------------------------------------

Consider a linear chain of $N$ coupled harmonic oscillators whose positions are defined by the coordinates $q_n$ with $n = 1,...,N$, respectively. In classical mechanics this system's Hamiltonian is
\begin{equation}\label{EqClassicalH}
H = \sum_{n=1}^N \left(\frac{p_n^2}{2m} + \frac{1}{2} \kappa q_n^2 + \frac{1}{2} \gamma (q_n - q_{n+1})^2\right),
\end{equation} 
where $p_n = m \dot q_n$. $\kappa > 0$ determines the strength of binding between the oscillating masses and their respective neutral positions, while $\gamma > 0$ controls the coupling between neighbours. For reasons of simplicity we impose periodic boundary conditions, i.e. $q_0$ is the same as $q_N$ and $q_{N+1}$ means just $q_1$. This system is the one-dimensional version of the `mattress' that Zee\cite{Zee} uses to introduce quantum field theory, or it can be seen as a simple model for phonons traveling through a crystal lattice.

We shall now follow the standard procedure: We will decompose the system's dynamics into normal modes which can be excited and evolve in time independently of one another. Mathematically speaking, this means to diagonalize the Hamiltonian $H$ via a Fourier transformation, revealing the system's equivalence to $N$ decoupled harmonic oscillators. One option is to expand $\vec q$ in real-valued sine and cosine modes, like in the article of  Johnson and Gutierrez\cite{JohnsonGutierrez}. Alternatively, a complex-valued decomposition into modes of $e^{ikn}$ can be applied like in the book of Greiner\cite{Greiner} on p.18. We shall also follow the latter approach and we will exhibit the technical steps in detail in order to make this article as self-contained as possible. We will deviate slightly from Greiner's path since we also need to find (among other things) the explicit coordinate transformation between the harmonic chain and the decoupled oscillators (eq. \ref{EqQtildeq} below).

We write the Hamiltonian (\ref{EqClassicalH}) as 
\begin{equation}
H = \frac{|\vec{p}|^2}{2m} + \frac{1}{2} {\vec{q}}^T D \vec{q},
\end{equation}
where $D$ is an $N \times N$ matrix taking care of the terms with $\kappa$ and $\gamma$ in (\ref{EqClassicalH}):
\begin{equation}
D = 
\begin{pmatrix} 
\kappa+2\gamma & -\gamma & 0 & 0 & \dots & 0 & -\gamma\\ 
-\gamma & \kappa+2\gamma & -\gamma & 0 & \dots &0 & 0\\ 
0 & -\gamma & \kappa+2\gamma  & -\gamma & \dots &0 & 0\\
\dots  & \dots & \dots & \dots & \dots &\dots & \dots
\end{pmatrix}
\end{equation}
The matrix $D$ is a circulant, which means that each row vector is rotated one element to the right compared to the preceding row vector. In this sense $D$ is generated from the $N$-vector $(\kappa + 2\gamma, -\gamma, 0, 0, ..., 0 , -\gamma)$ which forms the top line of the matrix. Now remember that any circulant matrix can be diagonalized via a discrete Fourier transform\cite{Davis} which, in turn, means to expand the position coordinates $q_n$ with respect to the orthonormal basis $\{\vec{f^{(k)}}\}$ with 
\begin{equation} \label{EqONB}
f^{(k)}_n = \frac{1}{\sqrt N} e^{-\frac{2\pi i}{N}k n} \quad \textrm{for} \quad k = -\frac{(N-1)}{2}, \dots, \frac{(N-1)}{2}.
\end{equation}
The vectors  $\vec{f^{(k)}}$ are also known as the `normal modes' of the system. We have made the assumption that $N$ is an odd number in order to save us a cumbersome treatment of special cases. Since $N$ is considered to be a large number in all practically relevant cases, this is no real restriction.
We call the position coordinates in the new basis $Q_k$ and they are related to the old ones by
\begin{equation}\label{EqFourierDecomp}
q_n =  \sum_k Q_k f^{(k)}_n = \frac{1}{\sqrt N} \sum_k Q_k e^{-\frac{2\pi i}{N}k n}.
\end{equation} 
Here and in the rest of the article a sum over $k$ always means that $k$ runs from $-(N-1)/2$ to $+(N-1)/2$ unless explicitly stated otherwise. Note that the $Q_k$ are complex numbers. In order to ensure that the coordinates $q_n$ remain real-valued, we have to impose the constraint
\begin{equation} \label{EqConstraint}
Q_k = \overline{Q_{-k}}.
\end{equation}
An analogous transformation maps the momentum $\vec p$ to the momentum $\vec P$ with $P_k = m \dot Q_k$. In the new coordinates the Hamiltonian takes the form
\begin{equation} \label{EqHnew}
H = \sum_k \left(\frac{|P_k|^2}{2m} + \frac{1}{2} \omega_k |Q_k|^2 \right).
\end{equation}
Here we have used the fact that our transformation was essentially a discrete Fourier transformation and it diagonalized the circulant matrix $D$. This implies that the eigenvectors of $D$ are the base vectors $\vec{f^{(k)}}$. We have chosen to call the respective eigenvalues $\omega_k$. Since $D$ is real symmetric, the $\omega_k$ are real. We have 
\begin{equation}
\omega_k \vec{f^{(k)}} = D \vec{f^{(k)}} = D \overline{\vec{f^{(-k)}}} = \overline{D \vec{f^{(-k)}}} = \overline{\omega_{-k} \vec{f^{(-k)}}} = \omega_{-k} \vec{f^{(k)}}
\end{equation}
and thus $\omega_k=\omega_{-k}$. In order to get rid of the complex coordinates before doing a canonical quantization, we perform a second coordinate transformation, splitting up each $Q_k$ into two real-valued coordinates:
\begin{equation} \label{EqQtilde}
Q_k = \frac 12 (1+i) \tilde Q_k + \frac 12 (1-i) \tilde Q_{-k} \quad\textrm{with } \tilde Q_k \in \R. 
\end{equation}
The constraint (\ref{EqConstraint}) is automatically observed then. By putting (\ref{EqQtilde}) into (\ref{EqFourierDecomp}) we get the transformation rule between the $q_n$ and the $\tilde Q_k$:
\begin{equation} \label{EqQtildeq}
q_n = \sum_k \tilde Q_k \frac{1}{2} \left((1+i) f^{(k)}_n+(1-i)f^{(-k)}_n\right) =: \sum_k \tilde Q_k \tilde f^{(k)}_n 
\end{equation}
We note that the $\{\tilde f^{(k)}_n\}$, in which we have developed $\vec q$, form another orthonormal basis in $\R^N$. This can be verified by direct computation, using the fact that the $\{f^{(k)}_n\}$ are also orthonormal.

The system's Hamiltonian (\ref{EqHnew}) looks very similar in the new coordinates:\\
\begin{equation} \label{EqHnew2}
H = \sum_k \left(\frac{\tilde P_k^2}{2m} + \frac{1}{2} \omega_k \tilde Q_k^2\right),
\end{equation}
where $\tilde P_k = m  \partial/\partial_t \tilde Q_k$. $H$ is now separated into $N$ one-dimensional harmonic oscillators with the eigenfrequencies $\omega_k$. Next, we perform a canonical quantization on $H$, promoting the location and momentum coordinates to Hilbert space operators. As a result we get a set of $N$ decoupled quantum oscillators. Solving the harmonic oscillator is one of the most common introductory problems in quantum mechanics and it is presented in many textbooks\cite{Liboff}. It can be achieved either by solving the oscillator's differential (Schroedinger) equation, or in an algebraic way based on commutation relations. Following the latter approach, we introduce the usual creation/annihilation operators $a^\dagger_k$ and $a_k$, so that the quantum version of (\ref{EqHnew2}) becomes
\begin{equation} \label{EqHnew3}
\hat H = \sum_k \omega_k \left(a^\dagger_k a_k + \frac 12 \right).
\end{equation}
The excitations of the harmonic chain created by the $a_k^\dagger$ are also called `particles' since this is how they often manifest themselves in experiments.

%-------------------------------------------------------
\section{Connecting the dots} \label{ConnectingDots}
%-------------------------------------------------------

By now we have established all the machinery needed to advance to the main ideas of this article, the Monte Carlo visualization of bosonic quantum fields. But let's first step back  and review how our work so far is related to first and second quantization. This section is meant to remind ourselves of the `big picture' and provide some orientation to novices in the field. It can be omitted by the advanced reader. 

We had started with one quantum particle which was free to move in two dimensions $q_1$ and $q_2$ subject to an harmonic potential. As known from introductory quantum mechanics courses, the particle's wave function (or, more precisely, the square of its absolute value) tells us the probability density of finding the particle at a given point in space when performing a measurement. We have visualized such a wave function for the specific eigenstate $\Psi_{2,1}$ in Fig. \ref{2DHarmonic}. The coordinates $q_1$ and $q_2$ are linked to location operators $\hat q_1$ and $\hat q_2$ which in turn describe a quantum measurement of the particle's position  in the respective coordinate.

Then we have shown how to solve the dynamics of the harmonic oscillator chain, which can serve for at least two physical models:

On the one hand, we can think of each point in the chain as being an actual physical particle like the atoms in a crystal lattice. In this case $q_n$ is still the spacial coordinate of each mass point relative to its equilibrium position, just like in the two dimensional case before, and $n$ simply enumerates the atoms of the chain. We are still in the `first quantization world' and if we were to measure the exact positions of the $N$ particles at the same time, the wave function of the system would tell us the probability of finding the chain is a certain configuration. The $\hat q_n$ remain simple location operators which describe a position measurement of the $n$th particle. For example, if the system's normalized state is $\Psi$ and we measure the position of the third particle, then the expectation value of the result is $\langle \Psi, \hat q_3 \Psi \rangle$. The normal modes of oscillation in the chain, which we have identified in Section \ref{SecLinHarChain}, are called phonons - an important concept in solid state physics. 

On the other hand, the chain can serve as a model for a bosonic quantum field, which belongs to the realm of `second quantization'. In this case $n$ becomes the discrete version of a spacial coordinate and $q_n$ is the field strength at the point $n$. The wave function of the system would give us the probability to encounter the field in a given classical state if it were possible to measure the field strength at all points in space at the same time. The operators $\hat q_n$ are now called `field operators' and they correspond to a measurement of the field strength in the spacial point $n$. In quantum field theory one would typically proceed to the continuum limit, replacing $n$ by $\vec x \in \R^3$ and $\hat q_n$ by some $\hat\phi(\vec x)$, for example. So all we have is a field, but where are particles - the bosons - which we wanted to describe? It turns out that what we call `particles' are simply excitation modes of the quantum field. Such excitations can be limited to a region in space and they can travel through the field very similarly to how particles in first quantization can be more or less localized and move in space. But, in contrast to first quantization, excitations can also become stronger or weaker - which means that particles can be created or destroyed. 

In the following section we shall see some examples of such excitations, how they can be visualized, and how they correspond to physical particles.

%-------------------------------------------------------
\section{Visualization of the Quantum Harmonic Chain} \label{SecVisualizationOfChain}
%-------------------------------------------------------

\subsection{Visualization Method}

We have noted in Section \ref{SecLinHarChain} that the Hamiltonian of the quantum harmonic chain of length $N$ is equivalent to the $N$-dimensional quantum harmonic oscillator (\ref{EqHnew2}) up to a coordinate transformation. The latter has the convenient property of being separable, so that its energy eigenstates are simply products of the familiar one-dimensional harmonic oscillator states:
\begin{equation}
\Psi_{\{\nu_k\}} = \prod_{k}{H_{\nu_k}(\sqrt{m\omega_k}\,\tilde Q_k)e^{-m\omega_k\tilde Q_k^2/2}},
\end{equation}

where $H_\nu$ is the $\nu$-th order Hermite polynomial and $\nu_k \in \N_0$ for $k = -(N-1)/2,\dots,(N-1)/2$ are the quantum numbers. 
Any wave function $\Psi$ which is given in terms of the coordinates $\tilde Q_k$ can be transformed back to the original coordinates $q_k$ via (\ref{EqQtildeq}). Once $\Psi(\vec q)$ is known, we can apply the Monte Carlo visualization which we have introduced in section \ref{SecMonteCarlo}. Compared to our simple example in Fig. \ref{2DHarmonic}, the main difference is that we now have $N$ coordinates instead of only two. Therefore we have to adapt our visualization prescription as follows:
\begin{enumerate}
\item Choose points $\vec q$ randomly according to a uniform probability distribution in a rectangular window within $\R^N$.
\item Each point $\vec q$ is visualized as a polyline in a parallel axes plot and the value of $\Psi(\vec q)$ is represented by the color of the polyline.  
\end{enumerate}
The visualization in a parallel axes diagram has the advantage that each point $\vec q$ is represented by a polyline which can be intuitively associated with the corresponding state of a classical (i.e., non-quantum) oscillator chain.

In order to plot such visualizations we have developed a small program in the numerically oriented programming language Scilab\cite{Scilab}. The source code is available on request from the author of this article.

\subsection{Ground State}

\begin{figure}[h!]
\centering
\resizebox{0.99\textwidth}{!}{
  \BW{\includegraphics{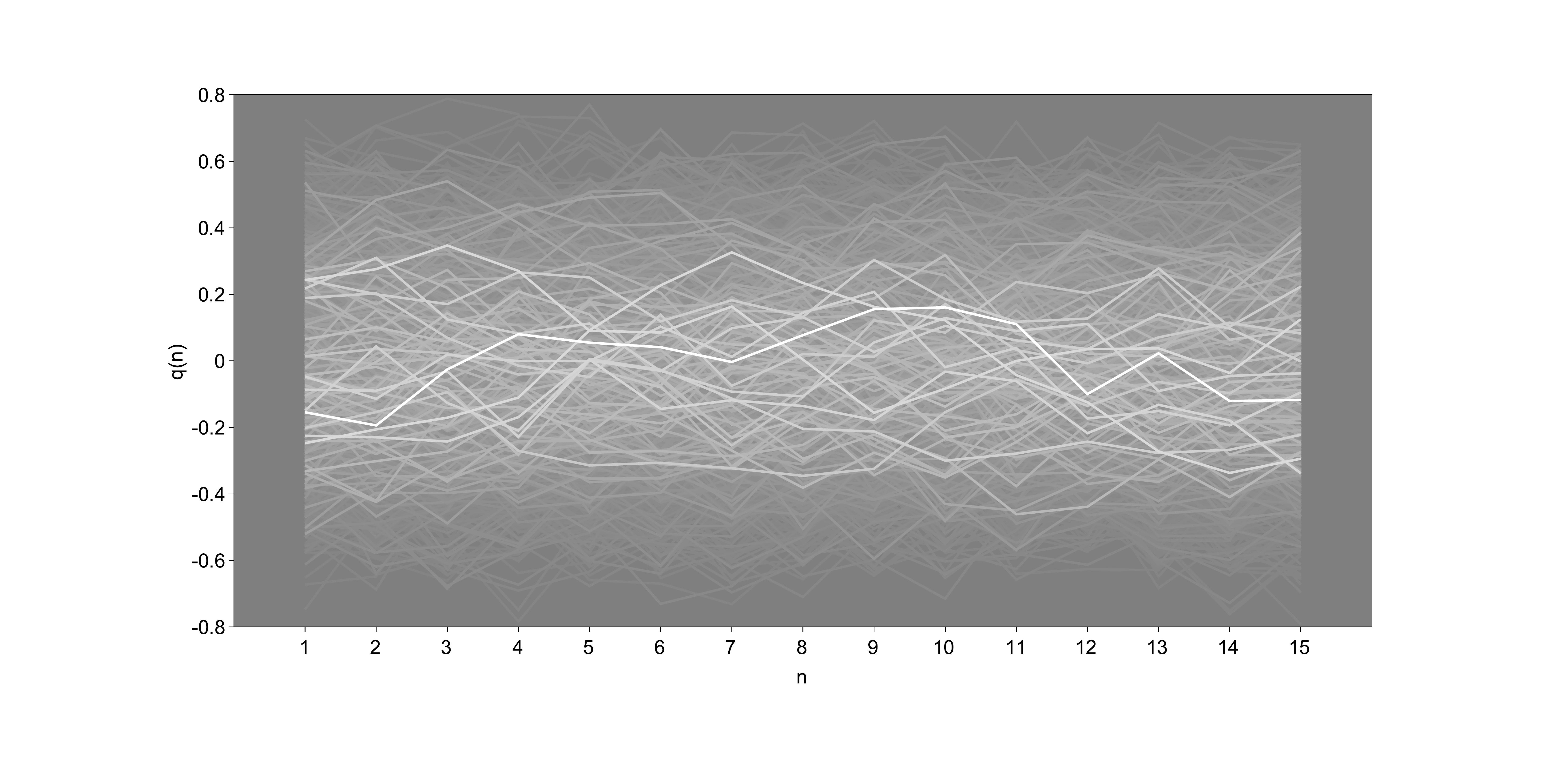}}
  \color{\includegraphics{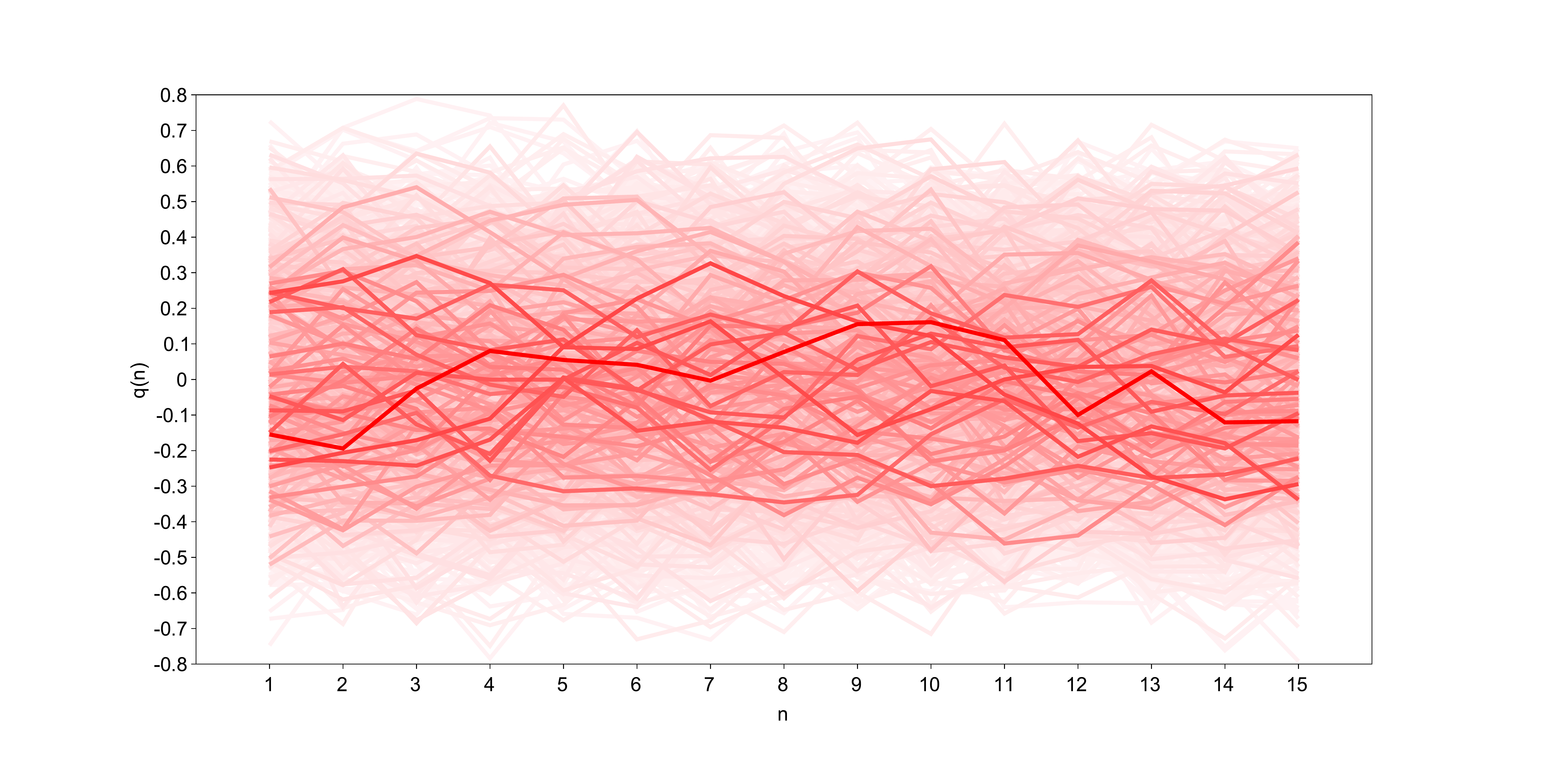}}
}
\caption{Monte Carlo representation of the ground state of a quantum harmonic chain. The wave function has been chosen real-valued and positive, thus all the lines are \BW{brighter than the background}\color{of the same color}. The \BW{brightest} lines \color{in deepest red }correspond to those $\vec q$ where the wave function assumes the highest values. Not surprisingly, in the ground state these are the lines close to $\vec q = 0$. Actually, a straight horizontal line corresponding exactly to $\vec q = 0$ would be drawn in the \BW{brightest}\color{deepest} color since it represents the global maximum of the ground state wave function. But due to the random nature of the Monte Carlo approach this line happens not to be drawn in the chart.}
\label{GroundState}
\end{figure}

As a first example of the results, Fig. \ref{GroundState} represents the ground state $\Psi_0 = \Psi_{\{\nu_k=0\,\forall k\}}$ of a quantum chain consisting of $N = 15$ oscillators. In the language of QFT, this state is called the `vacuum' since no excitations (=particles) are present. The wave function of the ground state is unique only up to multiplication with a complex number and in this case it has been chosen such that $\Psi_0$ is real and positive.  Each line in the plot represents one possible configuration of the 15 oscillators' positions. The \BW{brightness}\color{color} of each line corresponds to the value of the wave function $\Psi_0$ for the respective configuration. The color scale is chosen such that the value $\Psi_0 = 0$ blends in with the \BW{grey}\color{white} background and the \BW{brighter}\color{more colorful} the line, the higher the value of $\Psi_0$. Each polyline is plotted on top of the lines with lower $|\Psi_0|$, so that the most important lines (i.e. those with high $|\Psi_0|$) are more clearly visible in the chart. Those lines represent the most probable configurations of the chain in the sense of a quantum mechanical measurement and they tend to be relatively close to the $\vec q = 0$ line, which is the classical equilibrium state of the system. 

\subsection{Particles at rest}

\begin{figure}[h!]
\centering
\resizebox{0.99\textwidth}{!}{
  \BW{\includegraphics{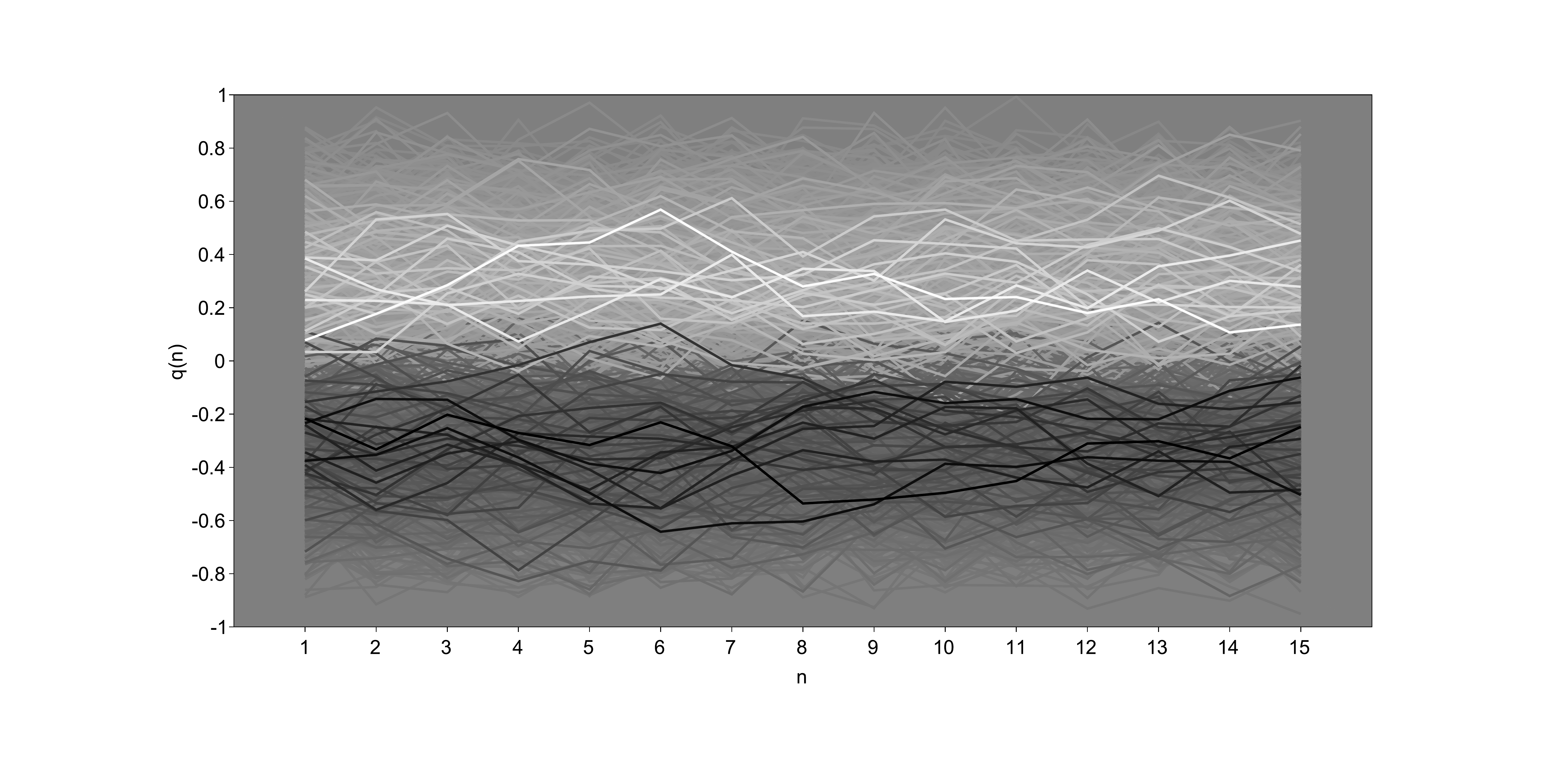}}
  \color{\includegraphics{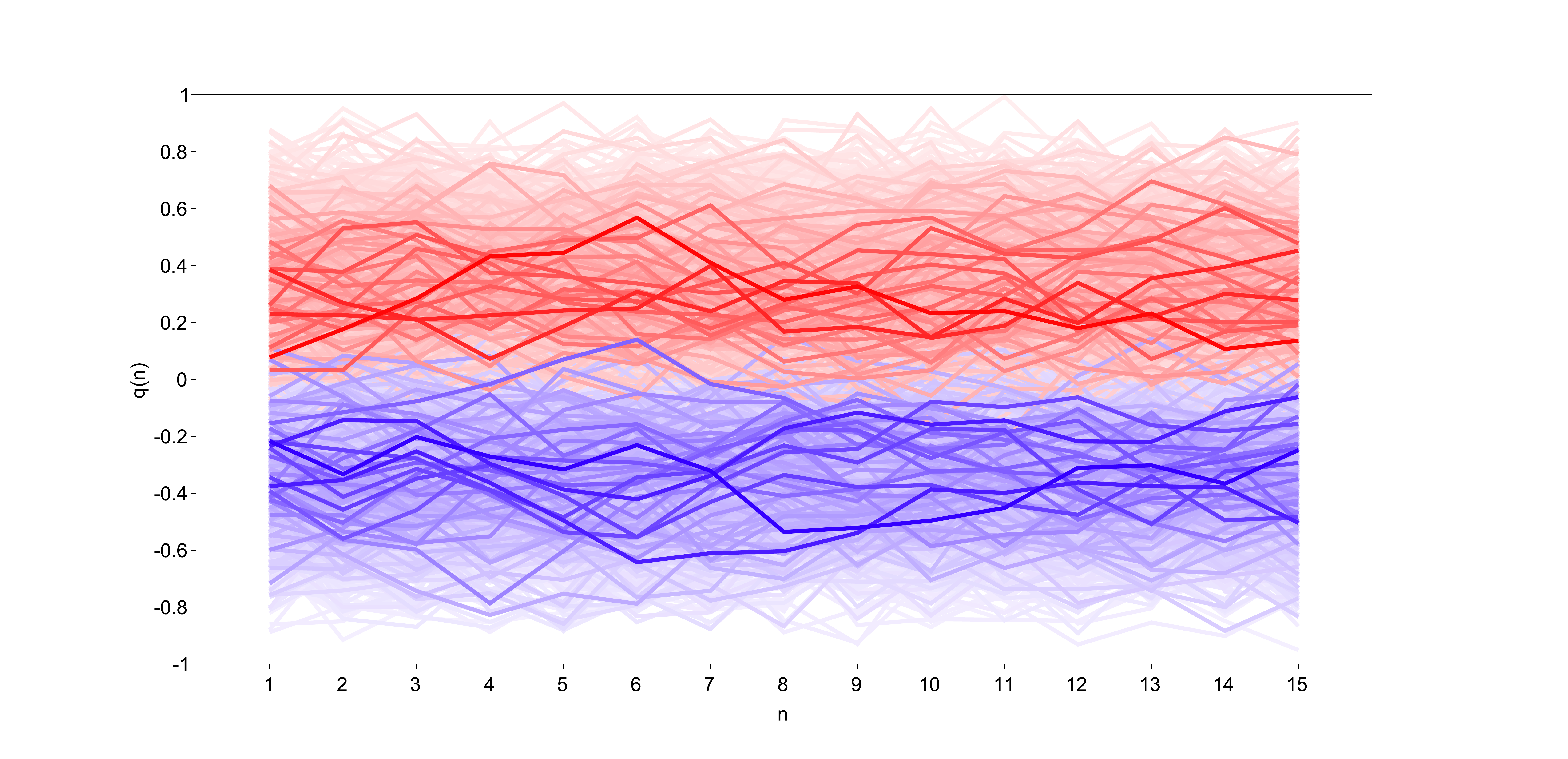}}
}
\caption{Monte Carlo representation of the first excited state of the harmonic chain with wave number zero. In the most prominent configurations of the chain, i.e., those corresponding to \BW{very bright}\color{deeply red} or \BW{very dark}\color{deeply blue} lines, the $q_n$'s tend to be either jointly positive or all negative. Thus, if we could perform a quantum measurement of the exact position of the harmonic chain, we would most likely either find it entirely shifted towards the positive $q$-direction or entirely the other way. In the particle interpretation of a quantum field this state would be called `one particle at rest'.}
\label{OnePartRest}
\end{figure}

\begin{figure}[h!]
\centering
\resizebox{0.99\textwidth}{!}{
  \BW{\includegraphics{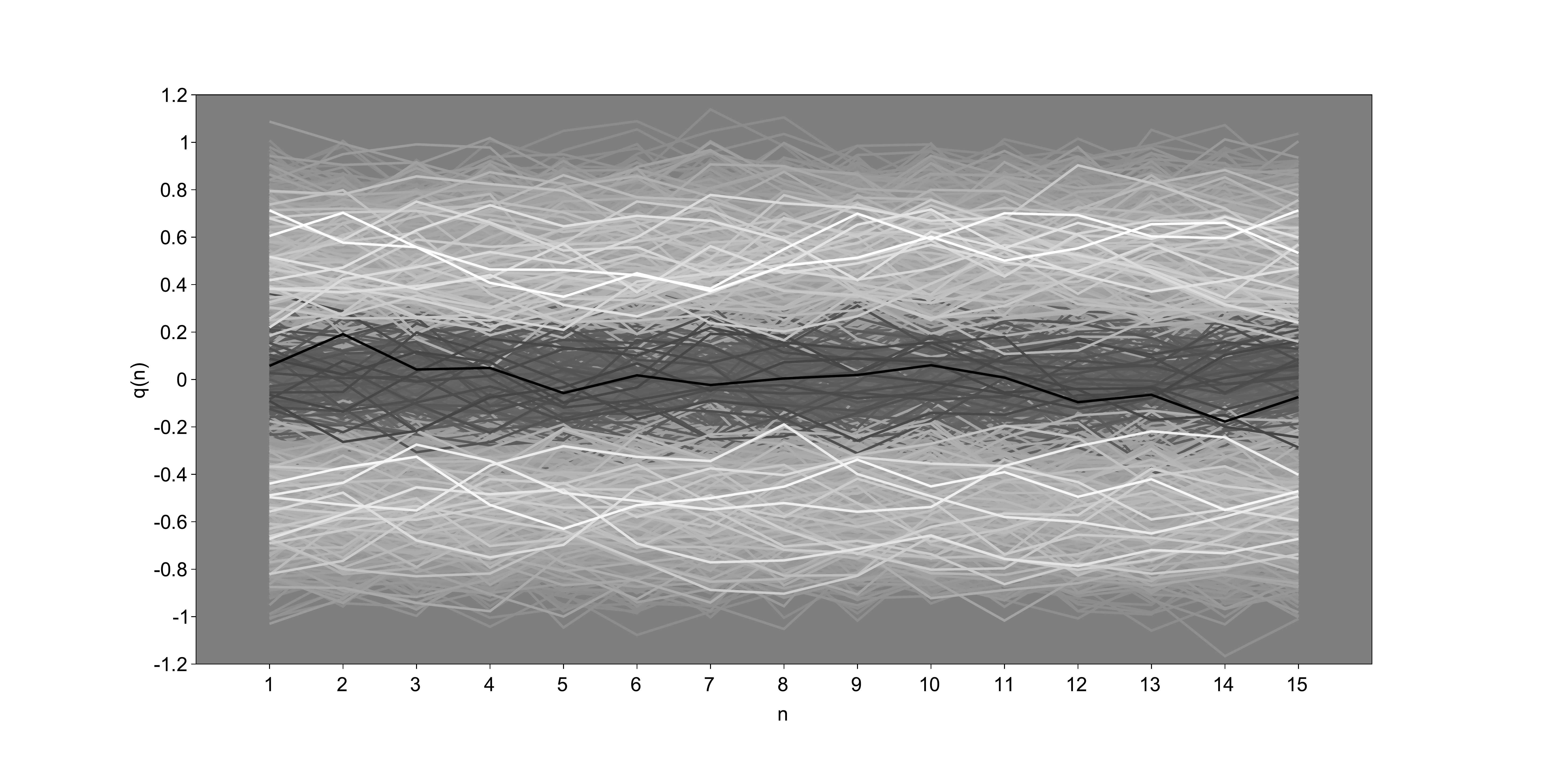}}
  \color{\includegraphics{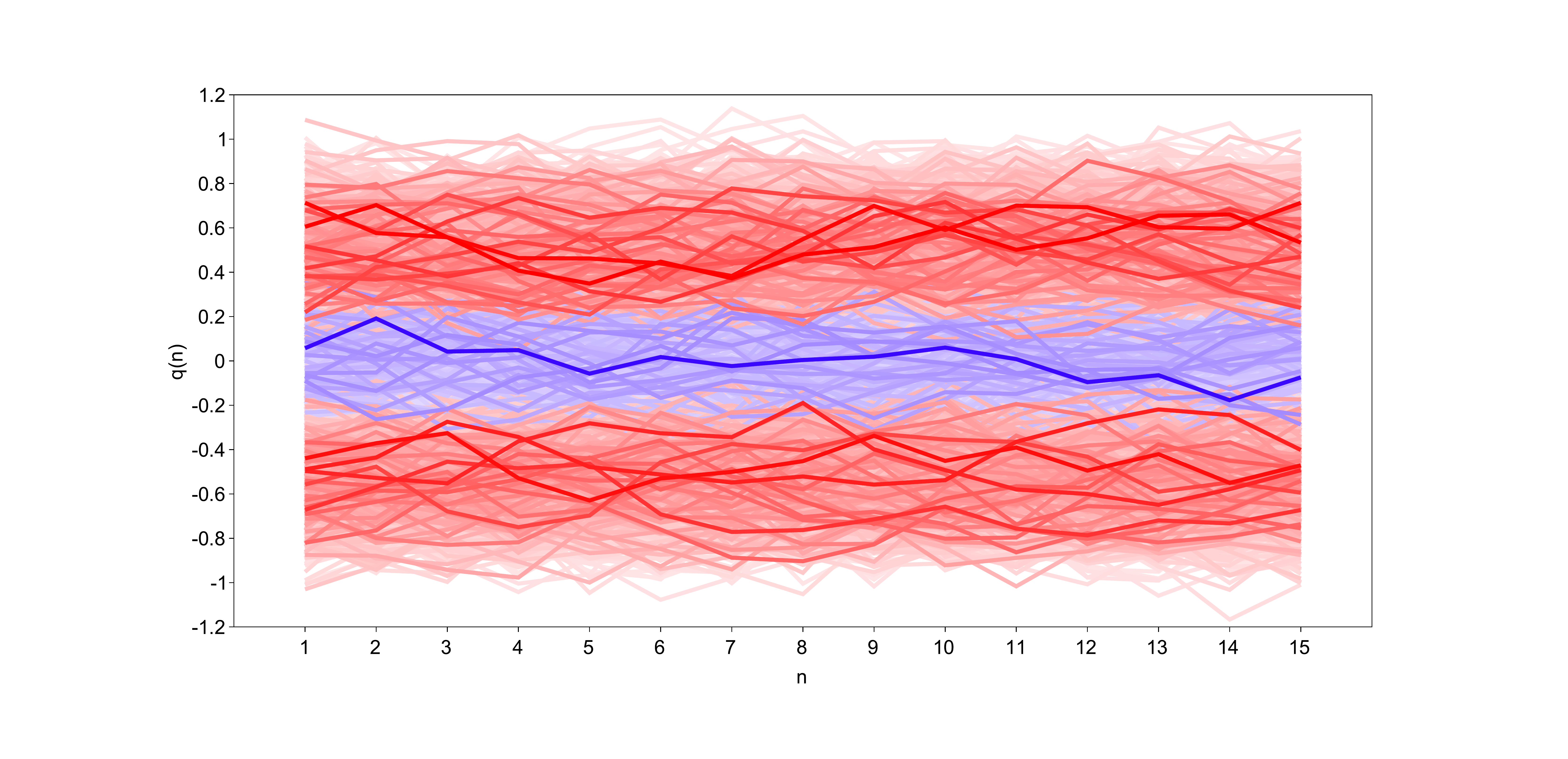}}
}
\caption{The second excited state of the harmonic chain with wave number zero is similar to the first one shown in Fig. \ref{OnePartRest}. But where the first excited state has only one `nodal line' at $\vec q = 0$ (or, much more precisely, one nodal hypersurface which contains the point $\vec q = 0$), we can now recognize two such `nodal lines' where the graph turns from being mostly \BW{bright}\color{red} to mostly \BW{dark}\color{blue} and back again.  In the particle interpretation of a quantum field, this state would be called `two particles at rest'.}
\label{TwoPartRest}
\end{figure}

As further examples we consider excitations of the chain which can be interpreted as one or two particles at rest, respectively. The one-particle state $\Psi = a_0^\dagger \Psi_0$ is visualized in Fig. \ref{OnePartRest} and the two-particle state $\Psi = (a_0^\dagger)^2 \Psi_0$ in Fig. \ref{TwoPartRest}. As before, the color scale is chosen such that configurations with $\Psi = 0$ blend in invisibly with the \BW{grey}\color{white} background, and the \BW{brightest/darkest}lines \color{in deepest red/blue }correspond to the largest (positive) / smallest (negative) values of $\Psi$. Similar to the ground state, the configurations with the  highest absolute value of $\Psi$ tend to be those with little fluctuation along the $n$-axis. But in the one-particle case we note that the polylines with mostly positive (negative) $q_n$ tend to correspond to positive (negative) values of $\Psi$. In the two-particle case the polylines close to the $\vec q = 0$ line tend to correspond to negative values of $\Psi$ while those farther away from that line are rather associated with positive values of $\Psi$. This pattern obviously resembles the positive and negative parts of the energy eigenfunctions of a one-dimensional harmonic oscillator. The classical analog to Fig. \ref{OnePartRest} and Fig. \ref{TwoPartRest} is a chain whose points masses oscillate synchronously and a stronger excitation of the chain as a whole corresponds to a higher number of particles in the quantum field.

\subsection{Particle with Non-Zero Wave Number}

\begin{figure}[h!]
\centering
\resizebox{0.99\textwidth}{!}{
  \BW{\includegraphics{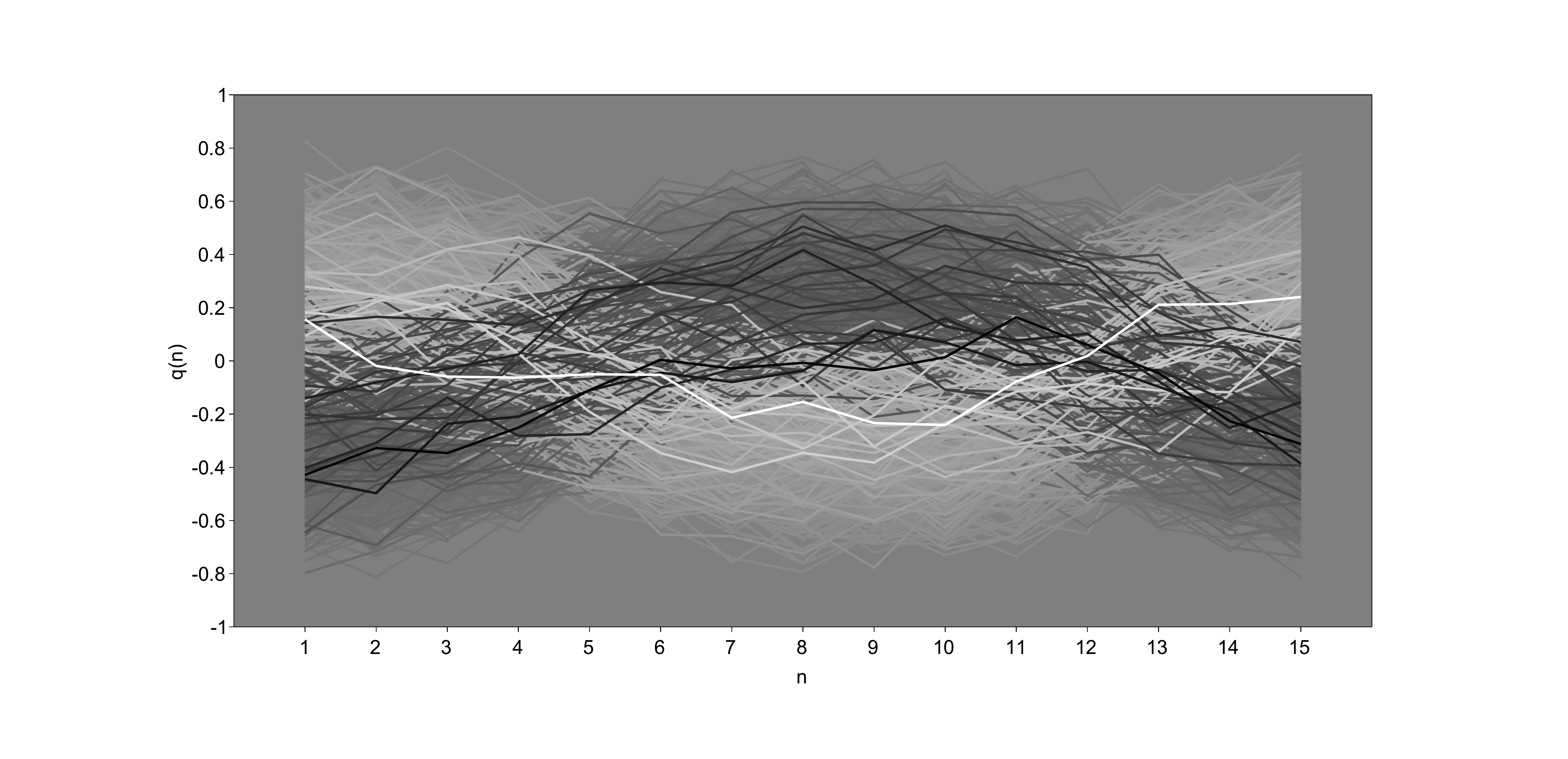}}
  \color{\includegraphics{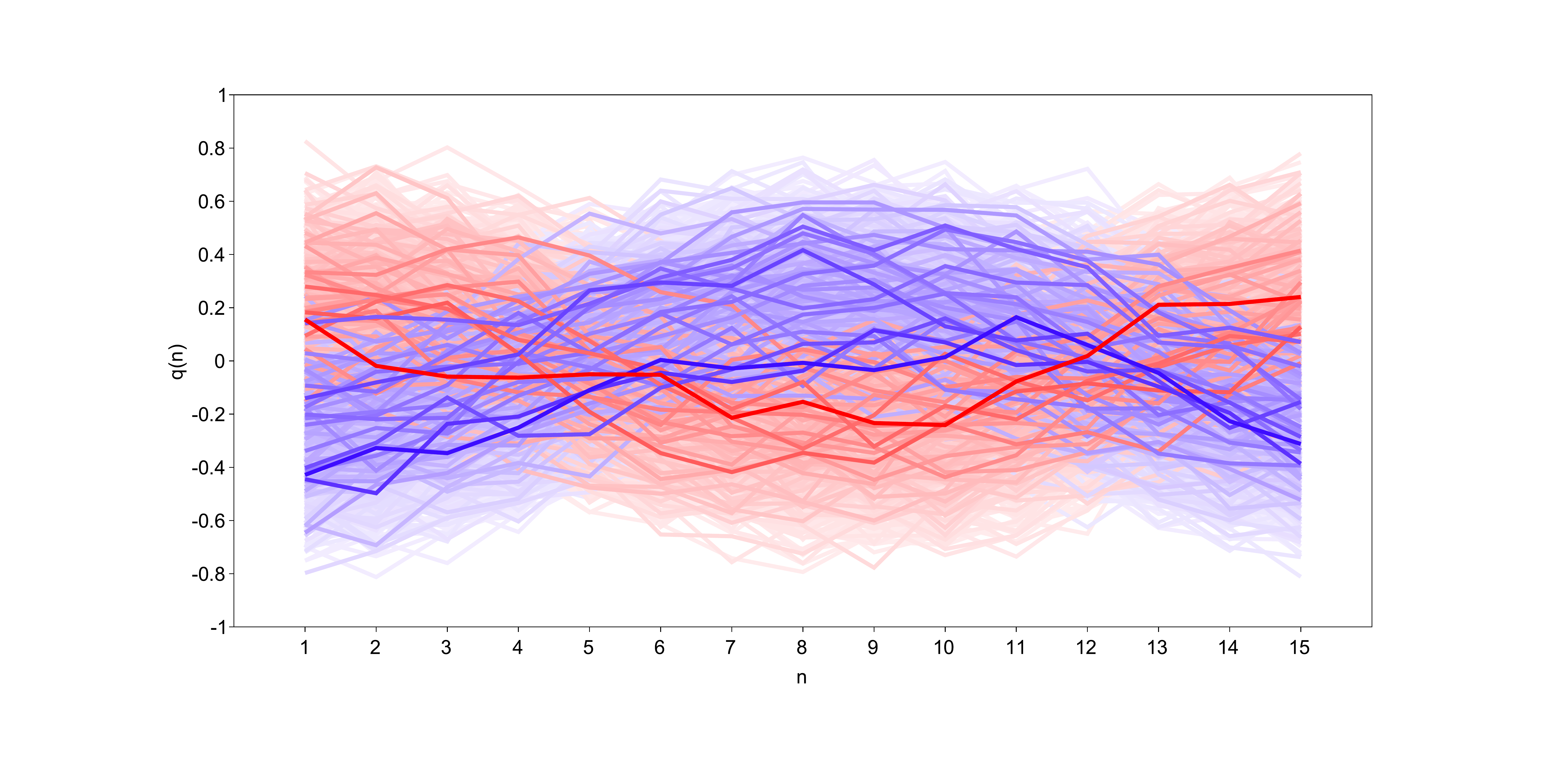}}
}
\caption{The first excited state of the harmonic chain with wave number $k=1$ corresponds to a standing wave with two nodes in classical mechanics. It can be interpreted as one quantum particle in an energy eigenstate of the lowest non-zero kinetic energy.}
\label{OnePartKOne}
\end{figure}

Now we turn to excitations of the chain with non-zero wave number. For example, the state $\Psi = a_1^\dagger \Psi_0$ (which we can also write as $\Psi_{\{\nu_k\}}$ where $\nu_k$ is equal to $1$ for $k=1$ and zero else) is represented in Fig. \ref{OnePartKOne} for $N=15$. Its classical analog is a standing wave in the oscillator chain. The location of the nodes near $n = 6$ and $n = 13$ can be explained by transforming back to the $q_n$ coordinates: In our state $a_1^\dagger \Psi_0$ the oscillating system is excited along the $\tilde Q_1$ coordinate, since this is how we defined the creation operators in (\ref{EqHnew3}). Using (\ref{EqQtildeq}) we see that an oscillation along $\tilde Q_1$ transforms back to an oscillation along the vector $\vec q$ with 
\begin{equation}
q_n = \frac{1}{2}\left((1+i) f_n^{(1)} + (1-i) f_n^{(-1)}\right)= \frac{1}{\sqrt{N}}\left(\cos\frac{2\pi n}{N} + \sin \frac{2\pi n}{N}\right), 
\end{equation}
which is zero for $n = 3N/8 \approx 5.6$ or for $n = 7N/8 \approx 13.1$.

A graph of the state $a_{-1}^\dagger \Psi_0$ would look very similar to Fig. \ref{OnePartKOne} with the exception that the standing wave is shifted by an offset of $N/4$ along the $n$-axis compared to the state $a_1^\dagger \Psi_0$. In the particle language of excitations $a_1^\dagger \Psi_0$ and $a_{-1}^\dagger \Psi_0$ would both be one-particle states, which is clear from how it is constructed via one creation operator.

We can also construct the quantum analog to a progressive wave in the chain, for example the state $(a_1^\dagger + i a_{-1}^\dagger) \Psi_0$. But the resulting wave function would not be real-valued anymore. A complex-valued $\Psi$ can be visualized with our method by using different colors, but for the present article we limit ourselves to real-valued $\Psi$ functions which can be visualized \BW{in gray scales}\color{with two colors only}.

\subsection{Localized Particle}

\begin{figure}[h!]
\centering
\resizebox{0.99\textwidth}{!}{
  \BW{\includegraphics{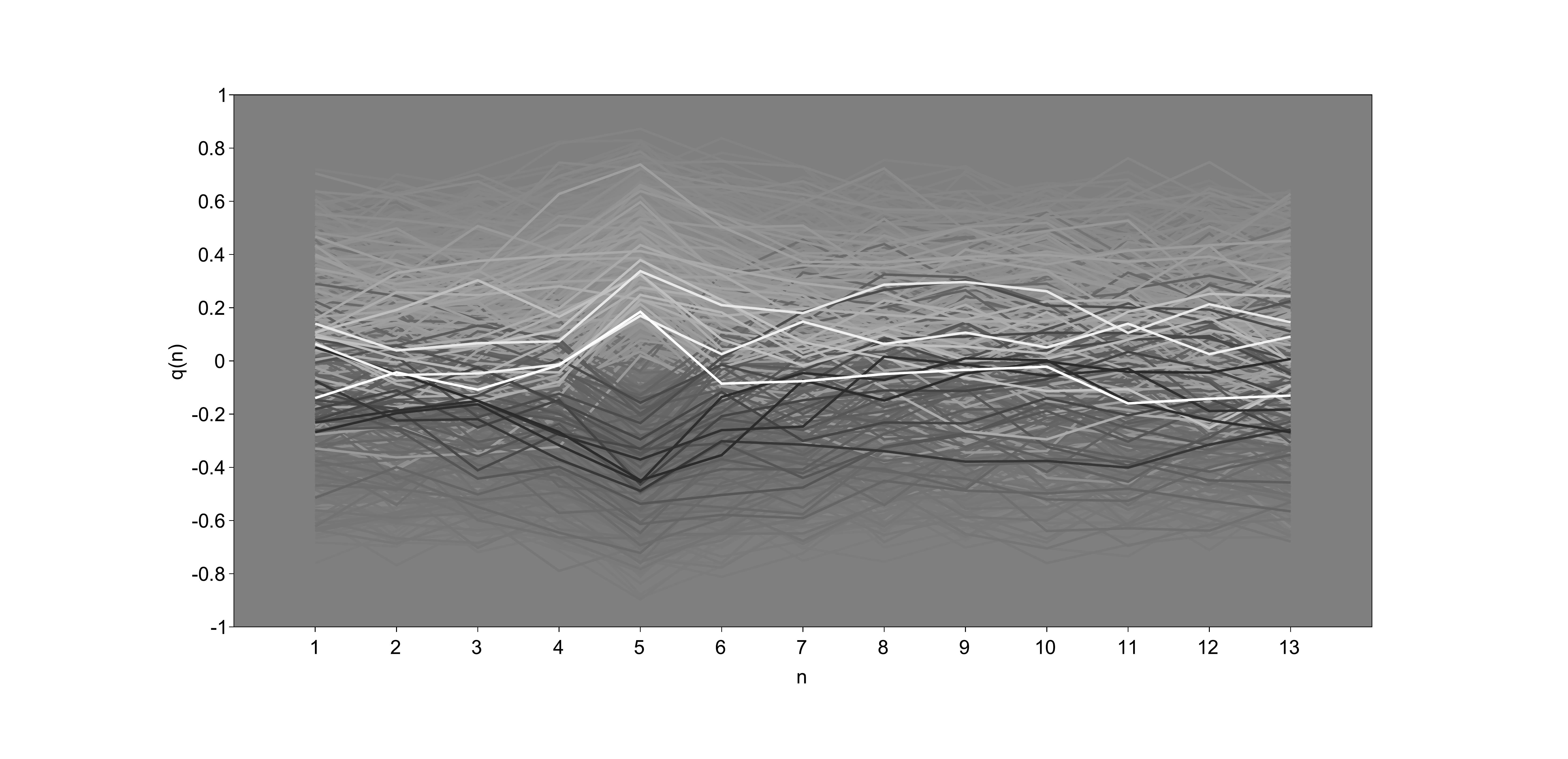}}
  \color{\includegraphics{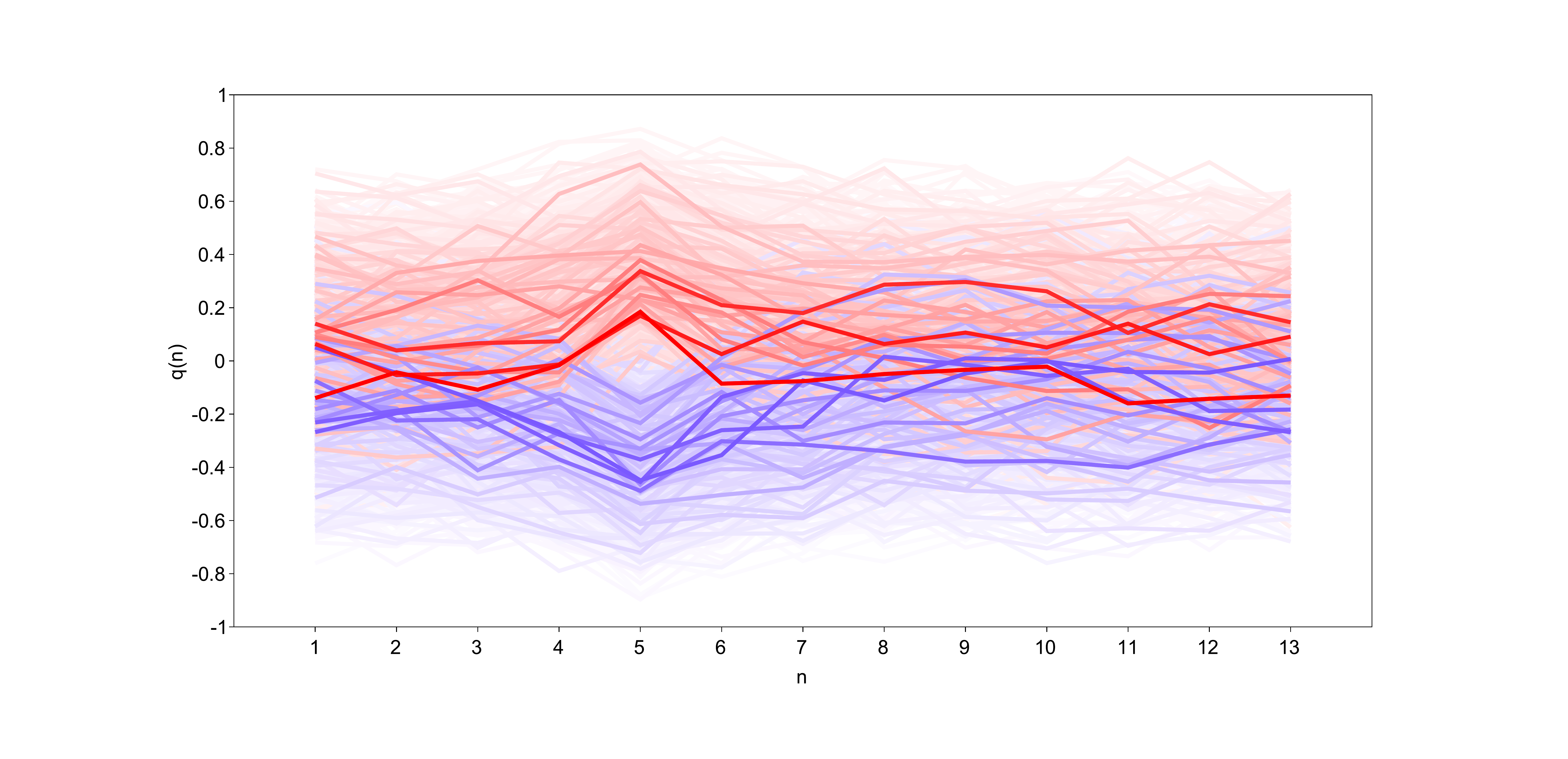}}
}
\caption{Monte Carlo plot of a local excitation corresponding to one localized particle at the position $n=5$. As opposed to the other states depicted so far, this state is not an energy eigenstate of the harmonic chain, which means that it is not essentially time-independent but would rather evolve into some spread-out oscillation pattern rapidly.}
\label{OnePartLocalized}
\end{figure}

Our next example is a state with one spatially localized particle. Recall that we had expressed $\vec q$ in the orthonormal basis $\{\tilde f^{(k)}_n\}$ in (\ref{EqQtildeq}) and note that the function $\delta_{n,n_0}$ (which is equal to one for $n=n_0$ and else zero) can be expanded in this basis as
\begin{equation} \label{EqDelta}
\delta_{n,n_0} = \sum_k \langle \tilde f^{(k)}_\cdot ,\delta_{\cdot,n_0} \rangle \tilde f^{(k)}_n =  \sum_k \tilde f^{(k)}_{n_0} \tilde f^{(k)}_n.
\end{equation}
 % = \sum_k \frac{1}{\sqrt{N}} e^{\frac{2\pi i}{N} k n_0} \frac{1}{\sqrt{N}} e^{-\frac{2\pi i}{N} k n}.
Remember that our operators $a^\dagger_k$ create a particle `in the state' $\tilde f^{(k)}_n$. Thus, in analogy to (\ref{EqDelta}) we construct new creation operators
\begin{equation} \label{EqCreatorsLocalized}
b^\dagger_n := \sum_k \tilde f^{(k)}_n a^\dagger_{k}
\end{equation}
and we expect $b^\dagger_n$ to create a particle localized in space at the position $n$. Fig. \ref{OnePartLocalized} is a visualization of the state
\begin{equation} \label{EqPsiLoc}
\Psi = b^\dagger_{5} \Psi_0.
\end{equation}
It is apparent that the sign of $\Psi(\vec q)$ depends only on the sign of the single coordinate $q_5$. This is exactly what one should expect from a quantum oscillator which has $N = 11$ (in the case of Fig. \ref{OnePartLocalized}) degrees of freedom, but which is excited only along the fifth coordinate. Yet the effect of the coupling between the point masses is also visible: At least at the positions $n=4$ and $n=6$ there is a notable correlation of the polylines' \BW{brightness}\color{color} with those at $n=5$ in the sense that \BW{white (black)}\color{red (blue)} lines dominate where $q > 0$ $(q<0)$. It looks like the excitation of the quantum field is `blurred' around that point. Where does this blur come from, given that $\Psi$ in (\ref{EqPsiLoc}) is constructed in close analogy to the $\delta$ function (\ref{EqDelta}) with its sharp peak? 
The deeper reason is that $a^\dagger_{k} \Psi_0$ is not only associated with a wave $\tilde f^{(k)}_n$, which stems from the Fourier transformation and which gives rise to the analogy between (\ref{EqDelta}) and (\ref{EqPsiLoc}), but also with an oscillation amplitude in the direction of the $\tilde Q_k$ coordinate. This amplitude depends on the strength of the harmonic potential in the $\tilde Q_k$ direction, which in turn depends (via the diagonalization of the system's Hamiltonian) on the coupling constant $\gamma$ between the mass points. Only in the trivial case $\gamma = 0$ the potential is radially symmetric in the space spanned by the $q_n$ or the $\tilde Q_k$ coordinates, which implies that the oscillation amplitudes of excited states don't depend on $n$ or $k$, so that the analogy between (\ref{EqDelta}) and (\ref{EqPsiLoc}) perfectly holds. In this case the `blurring' of the particle's position in Fig. \ref{OnePartLocalized} vanishes, which is also clear from the fact that for $\gamma = 0$ there is not even the concept of two `neighboring' point masses built into the physical system.

\subsection{Two Localized Particles}

\begin{figure}[h!]
\centering
\resizebox{0.99\textwidth}{!}{
  \BW{\includegraphics{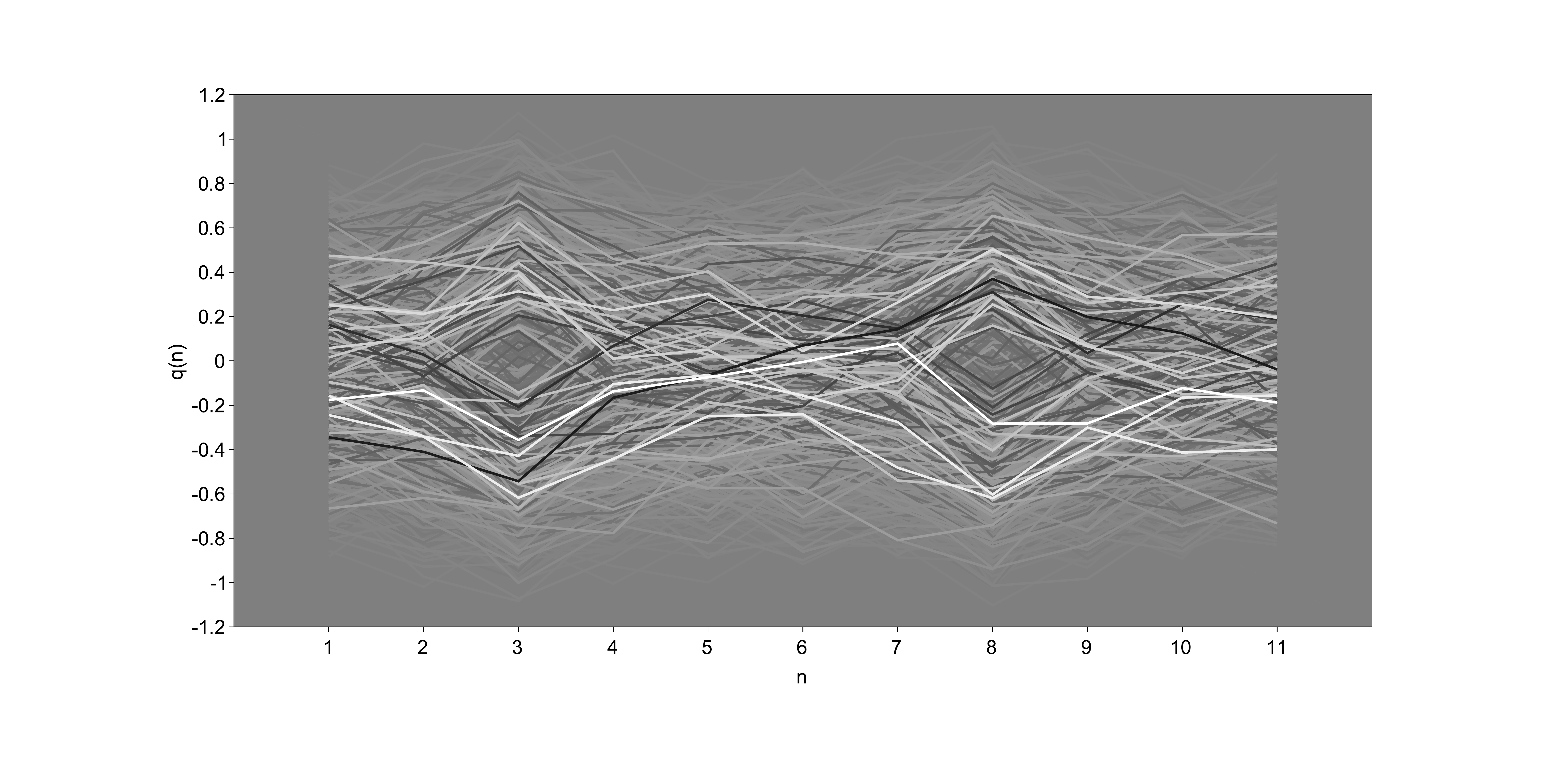}}
  \color{\includegraphics{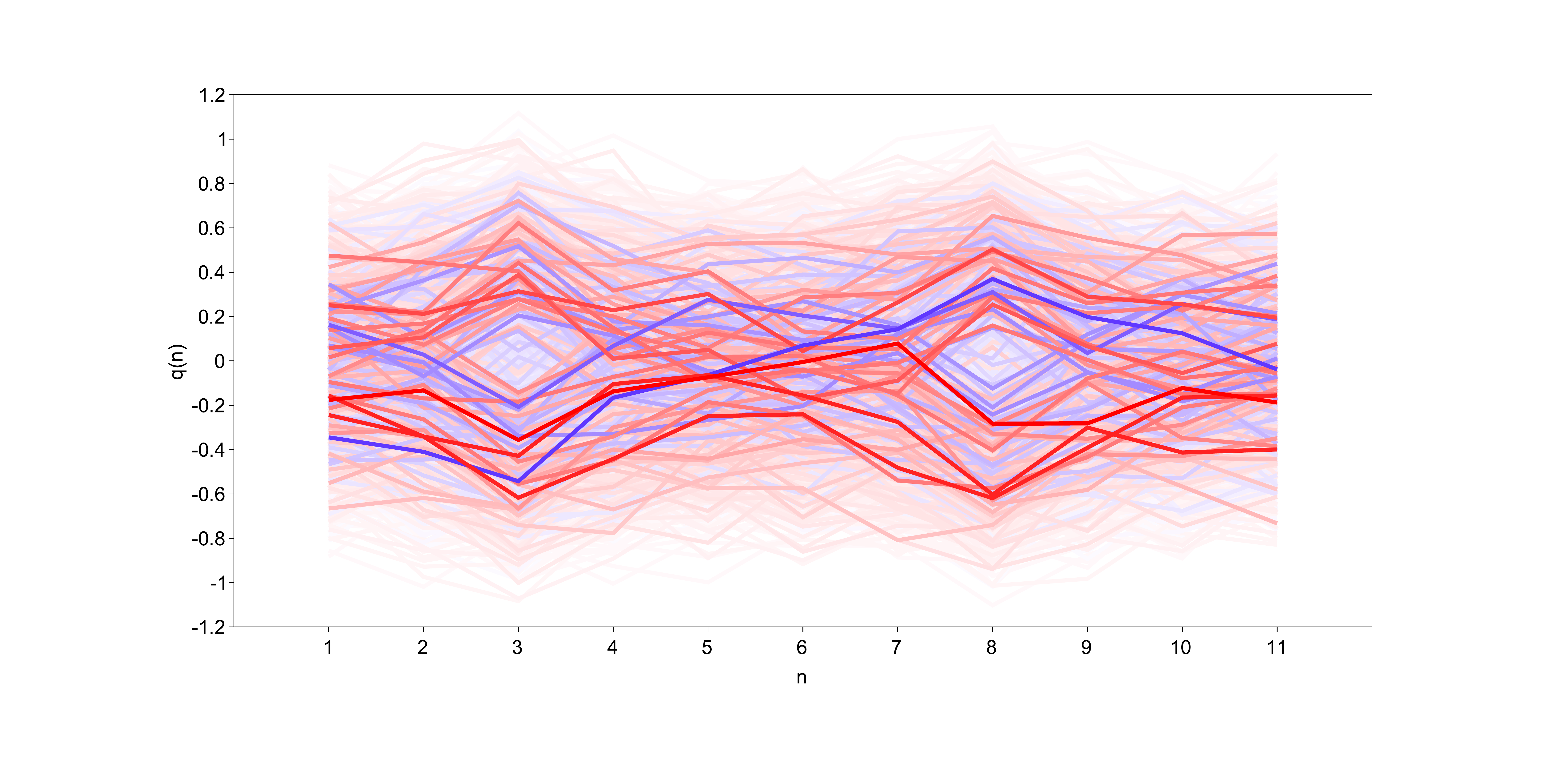}}
}
\caption{A state with two particles, one of them localized at $n=3$ and the other at $n=8$. While the peaks in excitation at those two locations are clearly visible, the detailed structure of the plot may seem chaotic at first glance. A closer inspection reveals a pattern in the \BW{white and black}\color{red and blue} lines which becomes much more evident in Fig. \ref{TwoPartLocalized-small-dist} below.}
\label{TwoPartLocalized-large-dist}
\end{figure}

\begin{figure}[h!]
\centering
\resizebox{0.99\textwidth}{!}{
  \BW{\includegraphics{TwoPartLocalized-small-dist-BW-v2.eps}}
  \color{\includegraphics{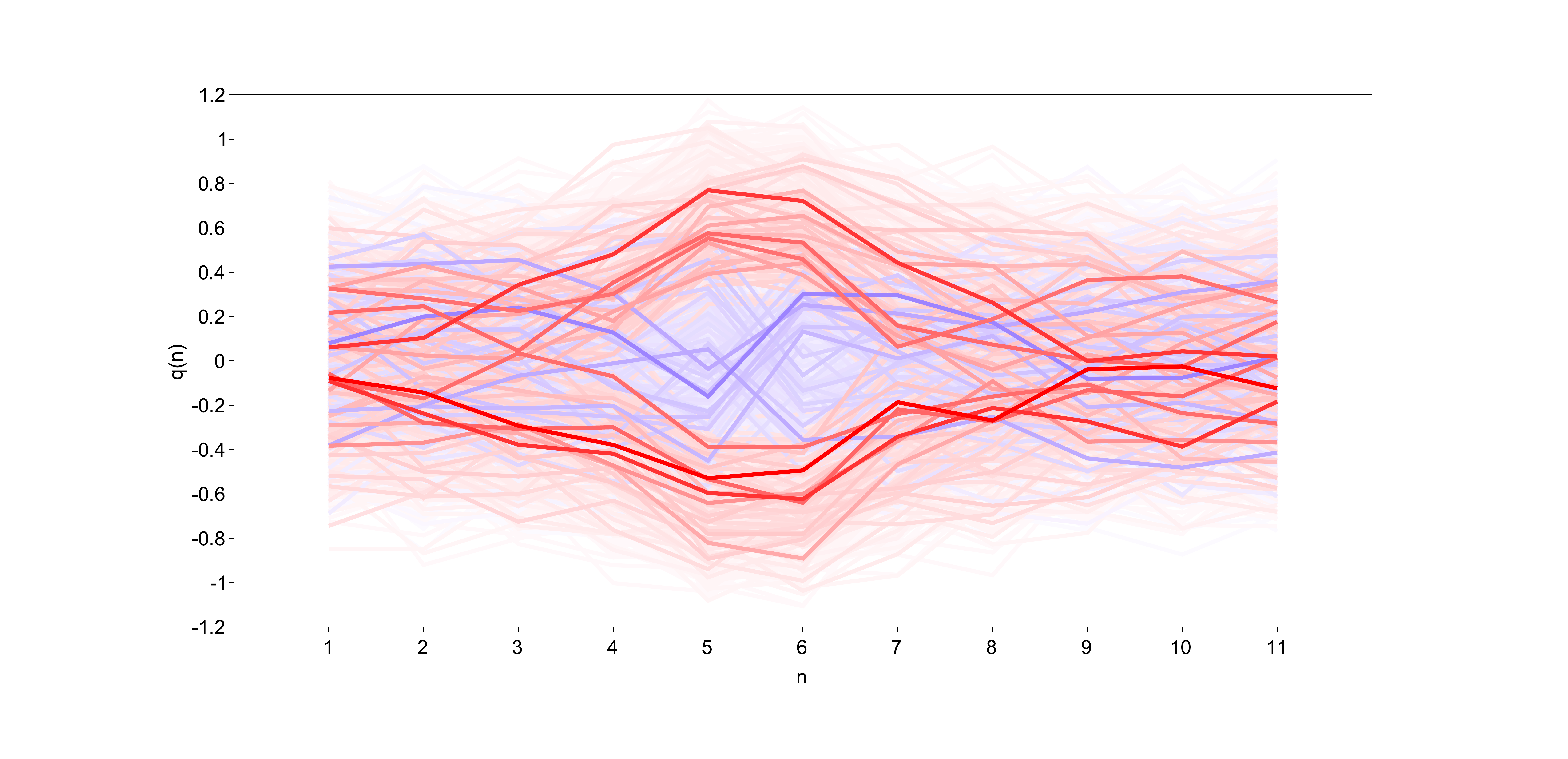}}
}
\caption{Two particles are localized very closely to each other at $n=5$ and $n=6$, respectively. The wave function assumes relatively large positive values for configurations of the harmonic chain where the atoms at $n=5$ and $n=6$ are strongly and synchronously displaced (i.e. where $q_5$ and $q_6$ are either both positive or both negative and of relatively large absolute value). It assumes negative values especially for configurations where the same two atoms are rather weakly and asynchronously displaced (i.e. $q_5$ is slightly positive and $q_6$ is slightly negative, or vice versa).}
\label{TwoPartLocalized-small-dist}
\end{figure}

\begin{figure}[h!]
\centering
\resizebox{0.99\textwidth}{!}{
  \BW{\includegraphics{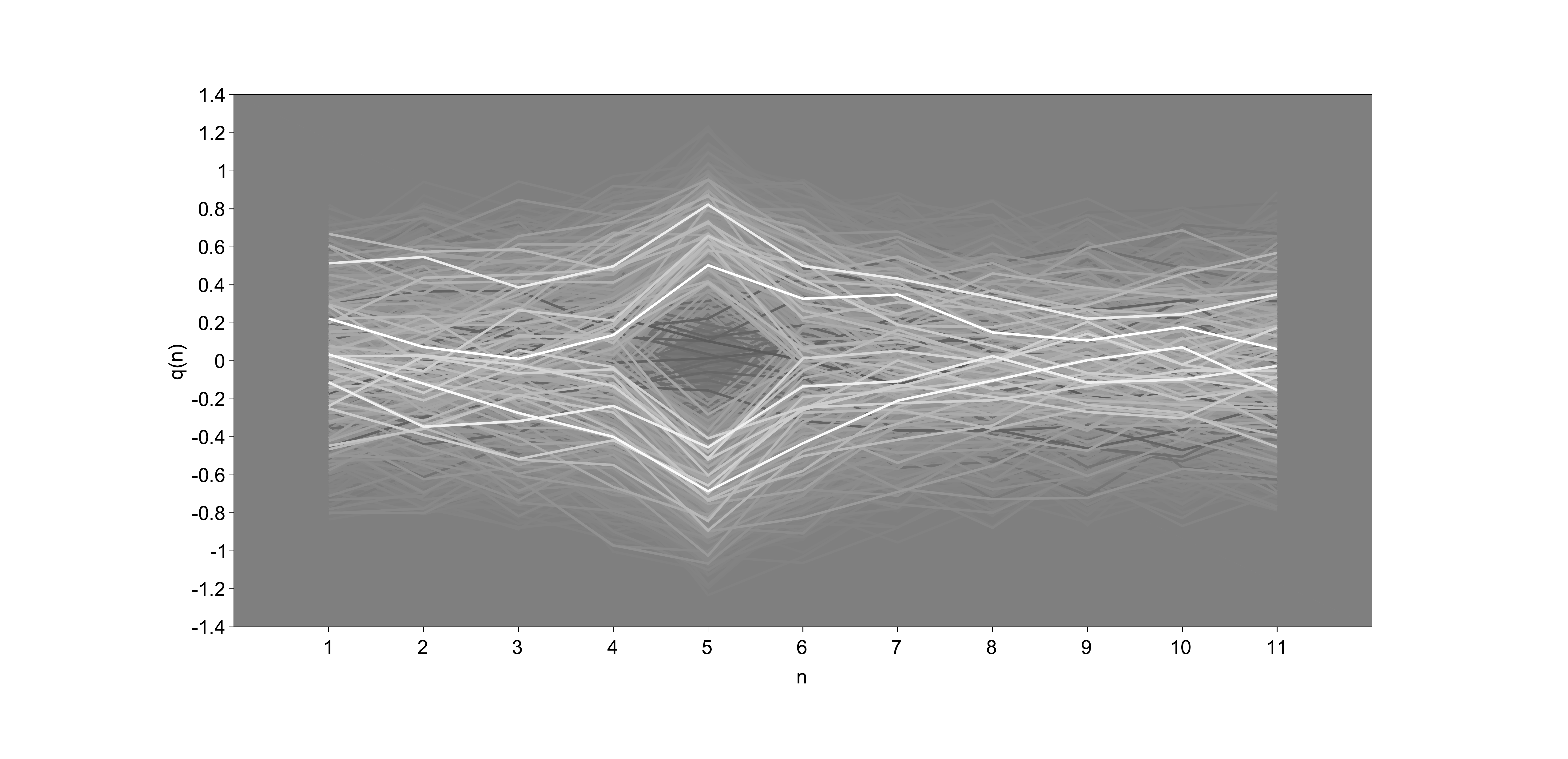}}
  \color{\includegraphics{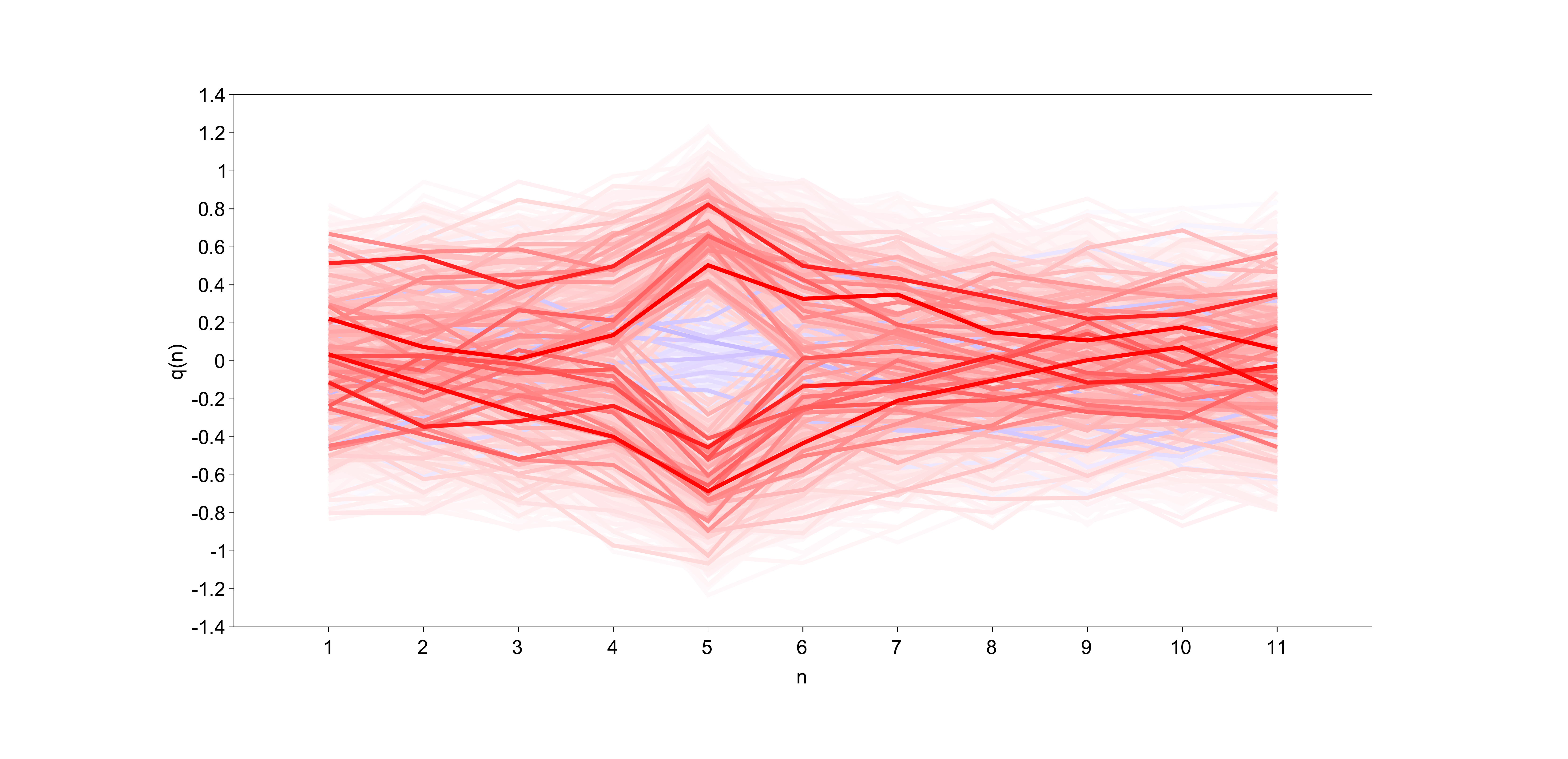}}
}
\caption{Two particles are in the same state, localized at $n=5$. The wave function $\Psi$ has two changes of sign along the $q_5$ axis: $\Psi$ is negative where $q_5 \approx 0$ and it is positive where the absolute value of $q_5$ is sufficiently large. This pattern resembles the second excited state of a one-dimensional quantum harmonic oscillator.}
\label{TwoPartLocalized-no-dist}
\end{figure}

In our final example we consider two localized particles, which are described by the state
\begin{equation} \label{EqPsi2Loc}
\Psi = b^\dagger_{n_1} b^\dagger_{n_2} \Psi_0.
\end{equation}
This case is particularly interesting because we can visualize how two bosons move gradually from two different states into one and the same. The Figures \ref{TwoPartLocalized-large-dist}, \ref{TwoPartLocalized-small-dist} and \ref{TwoPartLocalized-no-dist} contain Monte-Carlo visualizations for different values of the pair $(n_1,n_2)$, namely for $(3,8)$, $(5,6)$ and $(5,5)$, respectively. The two particles in Fig \ref{TwoPartLocalized-large-dist} are relatively far apart and each of the two localized excitations resemble the localized one-particle state from Fig. \ref{OnePartLocalized}, even though the phase structure of the $\Psi$ function (i.e. \BW{white}\color{red} lines vs. \BW{black}\color{blue} lines) is much more complicated. In Fig. \ref{TwoPartLocalized-no-dist}, on the other hands, two particles are located at the same point in space. Note the similarity of this plot with the two-particle state shown in Fig. \ref{TwoPartRest}: Both exhibit the pattern of an harmonic oscillator's second excited state (a function with two changes of sign), but in this case the excitation is limited to a small neighborhood of the point $n=5$. Fig. \ref{TwoPartLocalized-small-dist} shows how the transition between the two extreme cases (two particles far apart vs. localized in the same point) comes about.

%-------------------------------------------------------
\section{Discussion of Results} \label{SecDiscussion}
%-------------------------------------------------------

We have presented a new way of visualizing states of a quantum field and we conclude this article with a discussion of advantages and limitations of the new method, as well as an outlook for future investigation.

We hope that our visualizations can help students of QFT and solid state physics to develop a better intuitive understanding of some of their fundamental concepts. Compared to the method presented by Johnson and Gutierrez \cite{JohnsonGutierrez} we feel that our graphs are a slightly more direct representation of the quantum state, since they indicate the value of $\Psi$ for (a randomly chosen set of) points in the state space directly instead of showing a projection of $\Psi$ to the $q_n$ coordinates. For example, the spatial `blurriness' of a localized particle is clearly visible in Fig. \ref{OnePartLocalized} but not in the analogous Fig. 23 of Johnson's and Gutierrez' paper.

An inherent limitation of our method is that the graphs are not exactly reproducible. Due to the random nature of the Monte-Carlo approach two graphs generated with the exact same parameters will look similar, but not exactly the same. Also, the graphs are quite sensitive to parameters chosen in the Monte-Carlo algorithm. For example, if the number of random points chosen is too small, then none of these points might be in a region where $|\Psi|$ is large and the resulting graph will look quite noisy and without any clear structure. Whether that number is 'small' also depends on the number $N$ of point masses. The volume from which interesting points can be chosen (e.g. the set of all $\vec q \in \R^N$ with $|q_j| < 1$ for all $j = 1,\dots,N$) grows exponentially with $N$. Consequently, the necessary number of random points and the computation time of the algorithm also grow exponentially with $N$. Finally, for too `complex' states (e.g. those with more than just a few particles of different wave numbers) the graphs become too busy to be readable.  

In principle, our method could be extended to two-dimensional fields, where the polylines from the plots in the present paper would be replaced by overlapping surfaces in a 3D plot. In order to avoid a very busy and confusing graph, the number of surfaces drawn would have to be limited strictly to just a few with the very highest values of $|\Psi|$. 

An interesting subject for future work will be the extension of the method presented here to more complex phenomena in quantum fields, for example by considering interactions between fields or by making the shift from bosons to fermions.

%-------------------------------------------------------
\section{Acknowledgments}
%-------------------------------------------------------

The author is grateful to the anonymous referees for thorough reviews and several helpful suggestions which have made this article better.


\begin{thebibliography}{99}
\bibitem{Davis}
P.J. Davis: ``Circulant Matrices'' (Chelsea 1994)
\bibitem{DeAquinoEtAl}
V.M. de Aquino, V.C. Aguilera-Navarro, M. Goto, H. Iwamoto: ``Monte Carlo image representation'', Am. J. Phys., \textbf{69(7)} (2001)
\bibitem{Greiner}
R. Greiner: ``Field Quantization'' (Springer-Verlag Berlin Heidelberg 1996)
\bibitem{JohnsonGutierrez}
S.C. Johnson, T.D. Gutierrez: ``Visualizing the phonon wave function'', Am. J. Phys., \textbf{70(3)} (2002)
\bibitem{Liboff}
R.L. Liboff: ``Introductory Quantum Mechanics'' (Addison-Wesley Publishing Company, 1980)
\bibitem{LigareOliveri}
M. Ligare, R. Oliveri: ``The calculated photon: Visualization of a quantum field'', Am. J. Phys., \textbf{70(1)} (2002)
\bibitem{Thaller}
B. Thaller: ``Visual Quantum Mechanics: Selected Topics with Computer-Generated Animations of Quantum-Mechanical Phenomena'' (Springer/TELOS, Berlin, 2000)
\bibitem{Thaller2}
B. Thaller: ``Advanced Visual Quantum Mechanics'' (Springer New York 2004)
\bibitem{Zee}
A. Zee: ``Quantum Theory in a Nutshell'' (Princeton University Press, 2010)
\bibitem{Scilab}
Scilab Enterprises (2012). Scilab: Free and Open Source software for numerical computation (OS, Version 6.0.0) [Software]. Available from: http://www.scilab.org
\end{thebibliography}
\end{document}